\documentclass[oldversion]{aa}
\usepackage{epsfig}
\usepackage{graphics}
\usepackage{float}
\usepackage{amsmath}
\usepackage{multirow}
\usepackage{longtable}
\usepackage{rotate}
\usepackage{array}
\usepackage{subfigure}
\def\kms{\ifmmode{\rm km\,s^{-1}}\else\hbox{$\rm km\,s^{-1}$}\fi}

\setlongtables

\begin{document}

\title{Mass fluxes for hot stars}

\author{L.B.Lucy}
\offprints{L.B.Lucy}

\institute{Astrophysics Group, Blackett Laboratory, Imperial College 
London, Prince Consort Road, London SW7 2AZ, UK}
\date{Received ; Accepted }

\abstract{In an attempt to understand the extraordinarily small mass-loss rates
of late-type O dwarfs, mass fluxes in the relevant part of
($T_{eff}, g$)-space are derived from first principles
using a previously-described code for constructing
moving reversing layers. From these mass fluxes, a weak-wind domain is
identified
within which a star's rate of 
mass loss by a radiatively-driven wind is less than that due to
nuclear burning. The five weak-wind stars recently analysed by 
Marcolino
et al. (2009) fall within or at the edge of this domain. But although
the theoretical mass fluxes for these stars are $\approx 1.4$ dex lower than
those derived with the formula of Vink et al. (2000), the observed rates 
are still not matched, a failure that may reflect our poor understanding
of low-density supersonic outflows.\\
   Mass fluxes are also computed for two strong-wind O4 stars analysed by
Bouret et al. (2005). The predictions agree with the sharply reduced mass
loss rates found when Bouret et al. take wind clumping into account. 
}

\authorrunning{Lucy}
\titlerunning{Mass fluxes}
\maketitle
%________________________________________________________________

\section{Introduction}

Investigations of late-type O dwarfs in the SMC
(Bouret et al. 2003; Martins et al. 2004) and 
the Galaxy (Marcolino et al. 2009; M09) find that
mass-loss rates $\Phi$ determined by spectral modelling   
are at least an order of magnitude lower than those ($\Phi_{V}$) predicted 
with the formula of Vink et al. (2000).

Although this {\em weak wind problem} is considered to be a major problem
for stellar
wind theory (Puls, Vink, \& Najarro 2008), this may not in fact be so since the
$\Phi_{V}$'s
are semi-empirical estimates
based on Monte Carlo (MC) calulations of the deposition rates of
radiative energy in the supersonic winds. Accordingly, this conflict might 
be a consequence of the Vink
et al. assumptions concerning physical conditions in these winds. Moreover,
the global dynamical constraint imposed by Vink et al and previously by
Abbott \& Lucy (1985) does not guarantee that the derived $\Phi$'s are
consistent with stationary transonic flows.

In order, therefore, to make a more decisive confrontation between theory and 
observation, this paper reports an extensive grid of mass fluxes 
$J = \Phi/4 \pi R^{2}$ 
for O stars using a code described earlier (Lucy 2007; L07a). 
This code, which updates the moving reversing layer (RL)
theory of
Lucy \& Solomon (1970; LS70), uses a MC technique to estimate
the $J$ that allows the flow to accelerate continuously from sub- to
supersonic velocities. In other words, the eigenvalue $J$ is determined from
first principles by imposing a regularity condition at the sonic point.

\section{Solution technique}

Because the code is unchanged from L07a, details are omitted. However, the 
search procedure for the adjustable parameters has
been revised to facilitate a survey of $(T_{eff}, g)$-space.

\subsection{Implicit assumptions}

An unstated assumption in L07a is that the subsonic layers of a RL are
affected by the exterior supersonic layers {\em only} through 
backscattered radiation.
Specifically, therefore, no
structural changes occur due to information propagating back
into the RL via radiative-acoustic waves (Abbott 1980). This
assumption was checked in L07b and found to be justified: for the
models from L07a matching supersonic solutions were constructed and the
group velocity of Abbott waves computed. In each case, the direction
of information propagation is everywhere outwards towards higher velocities.

However, even without this justification, there are two further reasons for 
adopting this assumption. First, any conflict with observations
would then be possible evidence that structural changes do
indeed occur. Second, since such waves have to propagate back into the 
photosphere through an outflow that, according to most spectroscopists,
is chaotic and clumpy, their information content might well be scrambled and
lost.

A second unstated assumption is that after achieving supersonic velocities
matter continues to be accelerated at least until local escape
velocity is reached, so that no matter falls back onto the RL. 
Again, for the models of L07a, this is confirmed
by the matching exterior solutions reported in L07b.
But this assumption might well be violated for some late-type O stars
(Howk et al. 2000).

In addition, note that these dynamic RL's are
constructed {\em without} the Sobolev approximation, which, for a
microturbulent velocity of $~ 10 km s^{-1}$, is
not valid for Mach numbers $\la 3$ - see Fig.4 in L07a - and so is
inappropriate for the sub- and transonic flows of O stars.

\subsection{Adjustable parameters}

In addition to $J$, a model RL as described in L07a has two further adjustable
parameters $\delta$ and $s$. These define the velocity dependence
of $g_{\ell}$, the radiative acceleration due to lines, according to the
formula
\begin{equation}
   g_{\ell} = g_{*} max \left[\delta, \left(\frac{v}{a}\right)^{s} \right]
      \;\;\; with \;\;\; g_{*} = g - g_{e}
\end{equation}
where $g_{e}$ is the radiative acceleration due to electron scattering.  
With $0 < \delta < 1$ and $s > 0$, this formula is such that 
the effective gravity $ g_{eff} = g - g_{e} - g_{\ell} = 0$ at
$v = a$ and is negative at supersonic velocities, thus allowing the rapid
acceleration of the transonic flow to be accurately modelled - see Fig.2
in L07a.

\subsection{An identity}

If the equation of motion for plane-parallel, radiatively-driven isothermal
flow - Eq. (2) in L07a - is integrated between heights $x_{1}$ and 
$x_{2}$, we find that $Q_{1,2} \equiv 0$, 
where  
\begin{equation}
  Q_{1,2} = \frac{1}{2} (m_{2}^{2} - m_{1}^{2}) - \ell n (\frac{m_{2}}{m_{1}})
         + \frac{1}{a^{2}} \int_{x_{1}}^{x_{2}} g_{eff} dx
\end{equation}
Here $m = v/a$ is the flow's Mach number relative to the isothermal speed of
sound $a$.

In physical terms, $Q_{1,2} = 0$ implies that the gain in
mechanical energy between  $x_{1}$ and $x_{2}$ is
due to the work done by the gradients of gas and radiation pressures. 
But for a solution generated with the L07a code,
the identity will in general be
violated because of a non-optimum parameter vector $(J,\delta,s)$. 
Nevertheless, by
adjusting this vector as described below, we can find the solution giving
$Q_{1,2}=0$.

The above identity holds for any pair of points $(x_{1},x_{2})$. Here,
because of our interest in finding a solution representing smooth
transonic flow, we choose these points to be where $m_{1} \approx 0.5$ and    
$m_{2} \approx 2.0$, thus straddling the sonic point. Accordingly,
for this MC technique, the constraint $Q_{1,2}=0$ for the
interval $(m_{1},m_{2})$ is the analogue of the
regularity condition at $v = a$ in conventional investigations of transonic
flow.

\subsection{Determining the eigenvalue $J$}

The steps followed in obtaining a satisfactory model are:

1) For the chosen parameters $T_{eff}, g$, continuum fluxes are
extracted from the corresponding TLUSTY model (Lanz \& Hubeny 2003) at
26 frequencies $\nu_{k}(Hz)$ from 14.5-16.2 dex. This is illustrated in Fig. 1,
with 
parameters appropriate for a late-type O dwarf. In the MC calculation,
fluxes at the base of the RL are obtained by linear logarithmic
interpolation in the intervals $(\nu_{k}, \nu_{k+1})$. Note that because
the emergent MC flux at the top of the RL is constrained by Eq. (7) of
L07a to $= \sigma T_{eff}^{4}$, only {\em relative} continuum fluxes of the
TLUSTY models are used. 

\begin{figure}
\vspace{8.2cm}
\includegraphics{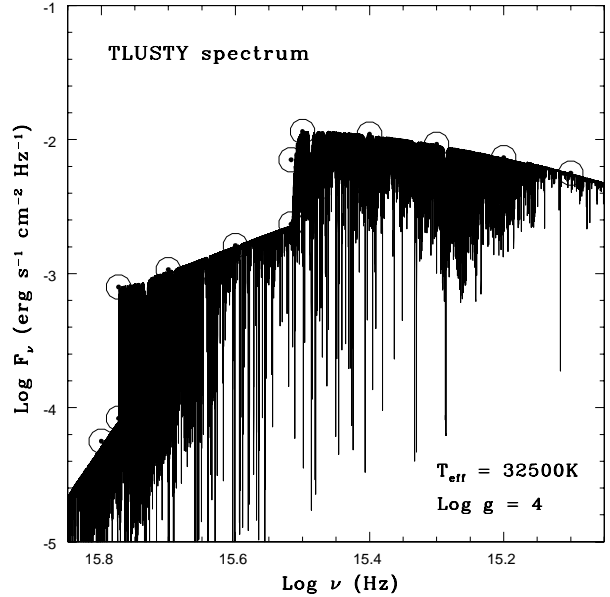}
\caption{Emergent spectrum of TLUSTY model atmosphere with parameters as
indicated. The open circles locate the
points used to approximate the continuum. }
\end{figure}

2) For the same TLUSTY model, ground-state departure coefficients are 
extracted from the point where $T_{e} \approx 0.75 T_{eff}$ in order to compute
ionization in the RL from Eq. (6) of L07a.

3) Initial values are selected for the parameters $(J,\delta,s)$, and the
resulting stratification of the RL computed as described in Sect. 2.3 of L07a. 
 
4) The radiation field throughout the RL is then derived with the MC
technique described in Sect. 3.2 of L07a. In particular, the MC estimator
$\tilde{g_{\ell}}$ for the radiative acceleration due to lines is evaluated.  
 
5) With these values $\tilde{g_{\ell}}$ , the integral in
Eq. (2) is approximated as a summation and $Q_{1,2}$ calculated.

6) Steps 3) to 6) are repeated for different $J$'s but fixed  $(\delta,s)$
in order to locate the intersection with $Q_{1,2} = 0$ - see Fig. 2.
The intersection is derived from a least squares linear
fit to 5-12 $Q$-values closely bracketing $Q = 0$.

\begin{figure}
\vspace{8.2cm}
\includegraphics{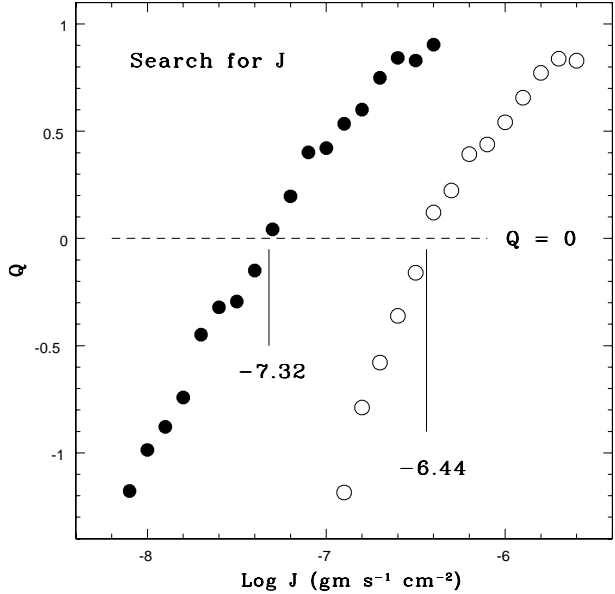}
\caption{The search for the eigenvalue $J$.
The atmospheres' parameters are $T_{eff} = 32500K$ (filled circles) and
$T_{eff} = 37500K$ (open circles) both with $log \: g = 4$. For each
atmosphere,
a sequence of models with varying $J$ but fixed $(\delta,s)$ is plotted.
The values of $log \: J$ given by the intersections with $Q_{1,2} = 0$ are
indicated.}
\end{figure}

7) From the model thus derived, fits to the variations of $\tilde{g_{\ell}}$
in the sub- and supersonic flows suggest improved values of $\delta$
and $s$, respectively. If these differ significantly from the current values,
steps 3) to 7) are repeated.

\subsection{Dynamical consistency}

A qualitative consistency check is the mismatch between 
$\tilde{g_{\ell}}$, the computed and $g_{\ell} (v; \delta,s)$,
the assumed radiative accelerations due to lines.
These are plotted for the model with 
$T_{eff} = 32,500K$ in Fig.3, which may be compared to the plot for 
$T_{eff} = 50,000K$ in L07a. This comparison shows that the adopted functional
form is here less successful at capturing the structure of the function 
$g_{\ell} (v)$. Clearly, a more flexible representation is
desirable to achieve better dynamical consistency.

\begin{figure}
\vspace{8.2cm}
\includegraphics{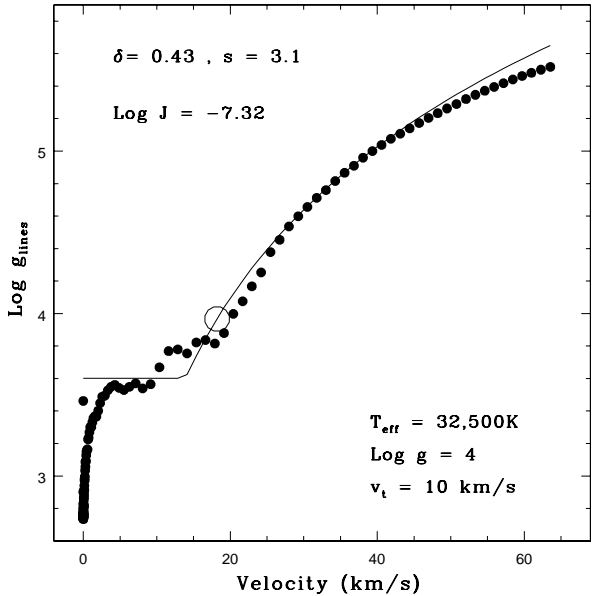}
\caption{Comparison of assumed and computed radiative accelerations due
to lines. The filled circles are the MC estimates and the solid line
shows the variation assumed in deriving the RL's stratification.
The open circle is the sonic point.}
\end{figure}

This lack of consistency implies some uncertainty in the predicted 
$J$'s. A rough estimate of this is derived as follows: 22 of the models in
Table 1 were originally computed with $\delta = 0.5$ and $s = 2$. When
$\delta$ and $s$ were optimized, the changes $|\Delta logJ|$ were $\leq 0.05$
in 13 cases and $ > 0.2$ in only 3 cases, with the largest difference
being $0.42$.

This experiment shows that the $J$'s are moderately insensitive to the vector
($\delta, s$) and thus to deviations from dynamical consistency. The most
important
step in deriving an accurate $J$ is locating the intersection  $Q_{1,2} = 0$, 
thus
ensuring that radiative driving makes its mandatory contribution to the
gain of mechanical energy as the flow accelerates through the sonic point
from $m_{1}$ to $m_{2}$. 

Given that the disagreement discussed in Sect. 1 is by at least 1.0 dex
in $\Phi$, the uncertainties in $J$ are not of present concern.

\section{Computed mass fluxes}  

In this section, $J$'s are calculated for parameter vectors $(T_{eff},g)$
chosen to illustrate the onset of powerful winds for O stars.
Results are given in Table 1 for 29 models with a range of
$g$'s appropriate for stars in their H-burning main sequence
phase. The models' composition is solar with $N_{He}/N_{H} = 0.1$
(Grevess \& Sauval 1998) and the included metal ions are as in Table 1 of
Lanz \& Hubeny (2003). The
microturbulent velocity $v_{t} = 10 km s^{-1}$. In cases where dynamical 
consistency is poor, $log J$ is followed by a colon. Note that two models
from L07a are recomputed in Table 1. The changes in 
$log J$ are +0.03 for D-50 and -0.07 for D-40. 

Throughout this paper, logarithmic values of $J$ are reported
with mass flux unit $gm \:s^{-1}cm^{-2}$. If the
radius of the star is known, its mass-loss rate $\Phi$ in solar masses
per year can then be derived from the formula
\begin{equation}
   log \: \Phi = log \: J + 2 \: log \: \frac{R}{R_{\sun}} + C
\end{equation}
with $C = -3.015$.

\begin{table}

\caption{Computed eigenvalues $J (gm \: s^{-1} cm^{-2})$}

\label{table:1}

\centering

\begin{tabular}{c c c c c}

\hline\hline

$T_{eff}(kK)$ &  $log \: g$  &  $\delta$ &  $s$  &  $log J$    \\

\hline
\hline

35.0  &  4.25   &  0.40   &  3.0  &  -7.32  \\

37.5  &         &  0.40   &  3.0  &  -6.68  \\

40.0  &         &  0.41   &  3.1  &  -6.27  \\

42.5  &         &  0.45   &  3.1  &  -5.93  \\

45.0  &         &  0.41   &  3.1  &  -5.78  \\

\cline{1-5}

27.5  &  4.00   &  0.26   &  2.7   &  -7.59  \\

30.0  &         &  0.45   &  2.7   &  -7.73:  \\

32.5  &         &  0.43   &  3.1   &  -7.32  \\

35.0  &         &  0.40   &  3.2   &  -6.97  \\

37.5  &         &  0.44   &  3.1   &  -6.43:  \\

40.0  &         &  0.47   &  3.1  &   -6.08:  \\

42.5  &         &  0.40   &  2.2  &   -5.64  \\

45.0  &         &  0.45   &  2.5  &   -5.39  \\

47.5  &         &  0.54   &  2.1  &   -4.93  \\

50.0  &         &  0.58   &  2.0  &   -4.38  \\

\cline{1-5}

27.5  &  3.75   &  0.17   &  3.3   &  -7.60  \\

30.0  &         &  0.44   &  3.1   &  -7.43  \\

32.5  &         &  0.47   &  3.3   &  -7.10  \\

35.0  &         &  0.48   &  3.6  &   -6.78:  \\

37.5  &         &  0.60   &  3.5  &   -6.37:  \\

40.0  &         &  0.60   &  2.5  &   -5.72  \\

\cline{1-5}

27.5  &  3.50   &  0.25   &  3.2  &   -7.56  \\

30.0  &         &  0.56   &  3.2  &   -7.33:  \\

32.5  &         &  0.47   &  1.6  &   -6.30  \\

35.0  &         &  0.55   &  2.0  &   -6.23:  \\

37.5  &         &  0.65   &  1.7  &   -4.74  \\

40.0  &         &  0.75   &  1.6  &   -4.20  \\

\cline{1-5}

27.5  &  3.25   &  0.44   &  3.6  &   -7.52:  \\

30.0  &         &  0.60   &  2.5  &   -6.64:  \\

\hline
\hline

\end{tabular}

\end{table}

\subsection{Weak-wind domain}

In thermal equilibrium, a star's mass loss rate via its wind equals that by 
nuclear burning when
$\Phi = L/c^{2}$, where $L = 4 \pi R^{2} \cal{F} $. Accordingly, we can
define the weak-wind domain in $(T_{eff},g)$-space to be where
$J < J^{*} \equiv \cal{F}$$/c^{2}$. The locus where $J = J^{*}$ has therefore
been derived by
interpolating between models in Table 1 and is shown as the bold line in
Fig. 4. Also shown is the contour where $J/J^{*}$ has increased to 0.5 dex,
illustrating the steepness of the onset of strong winds.

\begin{figure}
\vspace{8.2cm}
\includegraphics{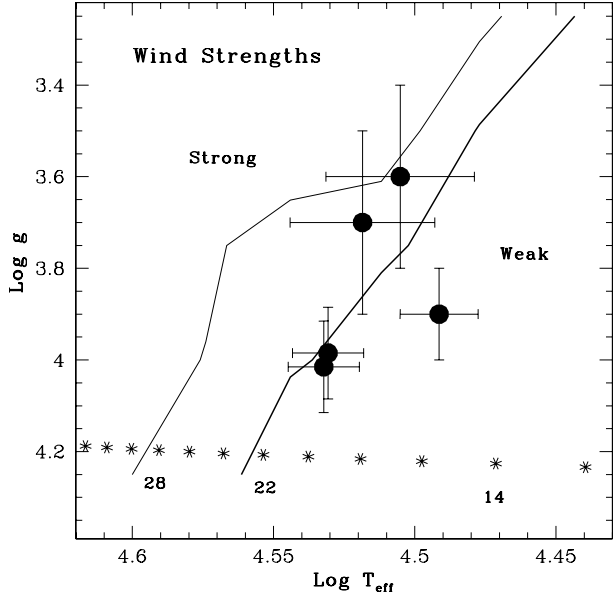}
\caption{Weak-wind domain. The bold line is the locus $J = J^{*}$ separating
weak from strong winds. Also shown is the locus where $J$ exceeds $J^{*}$
by 0.5 dex. The filled circles are the
weak-wind stars investigated in M09, and the
asterisks are ZAMS models of
Pols et al. (1998) for $Z = 0.02$ with masses 12(2)34 in solar units.}

\end{figure}

The five weak-wind stars analysed in M09 are
also plotted in Fig.4. They either fall within the weak-wind domain
or have error bars that just overlap the contour $J = J^{*}$.  

In order to determine the transition from weak to strong winds in terms
of stellar mass, zero-age main sequence (ZAMS) models from Pols et al. (1998)
for
$Z = 0.02$ are also included in Fig. 4. From this sequence, we see that stars  
on the ZAMS with $\cal{M/M_{\sun}} \la$ 22 are in the weak-wind domain while
stars with $\cal{M/M_{\sun}} \ga$ 28 are in the strong-wind domain.

The corresponding critical masses can be derived for the mass-loss formula of
Vink et al. (2000). From their Eq. (12) with $v_{\infty}/v_{esc} = 2.6$
(Lamers et al. 1995) and noting that the formula applies only down to
$T_{eff} = 27,500K$, we find that the upper limit for weak winds drops
from 22 to $<$ 12, while the lower limit for strong winds drops from 28 to
14 solar masses. These large changes have profound implications for the
evolution of massive stars.

\subsection{Individual Marcolino stars}

Finding that the stars of M09 fall in or at the edge of
the weak-wind domain in Fig. 4 appears to suggest that the $J$'s predicted from first
principles with the L07a code solves the {\em weak wind problem}. But before
accepting this conclusion, we must compare observed and predicted mass
fluxes for the five stars individually. This is done in Table 2, where
the data is as follows:

\begin{table}

\caption{Mass fluxes for observed stars}

\label{table:2}

\centering

\begin{tabular}{c c c c c }

\hline\hline

$Star$ &  $log J_{M}$  &  $log J_{L}$ &  $log J_{V}$  &  $log J^{*}$   \\

\hline
\hline

HD 216898    &  -8.0 $\pm$ 0.7   &  -7.1 $\pm$ 0.2   &  -5.7  &  -7.1  \\

HD 326329    &  -8.0 $\pm$ 0.7   &  -7.5 $\pm$ 0.2   &  -6.0  &  -7.2  \\

HD  66788    &  -7.8 $\pm$ 0.7   &  -7.1 $\pm$ 0.2   &  -5.7  &  -7.1  \\

$\zeta$ Oph  &  -7.7 $\pm$ 0.7   &  -6.8 $\pm$ 0.8   &  -5.7  &  -7.2  \\

HD 216532    &  -8.0 $\pm$ 0.7   &  -6.9 $\pm$ 0.7   &  -5.5  &  -7.1  \\

\hline
\hline 

\end{tabular}

\end{table}

Column 2: $\: log \: J_{M}$ 
computed from the measured $log \:\Phi$'s reported in M09 together with their
conservative error estimates. 

Column 3: $\: log J_{L}$ obtained by interpolation
in Table 1 for the $T_{eff}$ and $g$ given in Table 3 of M09. The computed
errors
reflect only the propagation of the errors in $T_{eff}$ and $g$ given in
M09 and shown in Fig. 4.

Column 4: $\: log J_{V}$ derived from Eq. (12) in Vink et al. (2000) with
parameters from Table 3 in M09. 

Column 5: $\:log J^{*}$, where $J^{*} \equiv {\cal{F}}/c^{2}$.

From Table 2, we see that the $J_{L}$'s are systematically smaller than
the $J_{V}$'s by on average 1.4 dex. Nevertheless, this huge reduction is
still not enough to explain the measured values $J_{M}$, which
are smaller than the $J_{L}$'s by on average 0.8 dex. 

Given the large error bars, the individual differences 
$log J_{M} - log J_{L}$ are barely significant. But with all differences being
of the same sign, the data as a whole could be said to reject the hypothesis
that the L07a theory correctly predicts observed $J$'s.
But this is not purely a question of statistics. The reliability of
the diagnostic investigations can be questioned, especially for low $\Phi$'s
when
often only the C IV doublet is available for analysis, and this line's weak or
absent emission component is unexplained (Martins et al. 2004). This
latter problem emphasizes how limited is our understanding of the supersonic
zones of these stars' low density winds, a problem affecting the
Vink et al. (2000) predictions as well as diagnostic codes.

\subsection{The matching problem}

As noted in Sect.1, $\Phi$'s derived from a global dynamical constraint
may be inconsistent with stationary transonic flow; and the same remark
applies to $\Phi$'s derived diagnostically from circumstellar spectra.
To illustrate the
magnitude of these inconsistencies for the weak-wind stars, the 
consistency check carried out in Fig.3 has been repeated in Fig.5 
with $J_{L}$
increased by 1.4 dex to represent the Vink et al recipe and decreased
by 0.8 dex to represent the Marcolino et al. estimates. In both cases,
there are huge differences between $g_{\ell} (v)$ and
$\tilde{g_{\ell}}$, indicating gross violations of dynamical consistency.
Moreover, searches in ($\delta, s$)-space with
$J$ fixed at these high or low $J$'s fail to find alternative 
dynamically-consistent solutions. 

Evidently, any estimate of $\Phi$ derived from an analysis of a wind's
supersonic flow should ideally be confirmed by
matching to a stationary transonic flow, thus exhibiting a successful
transition to the quasi-static atmosphere where the star's absorption line
spectrum is formed.

\begin{figure}
\vspace{8.2cm}
\includegraphics{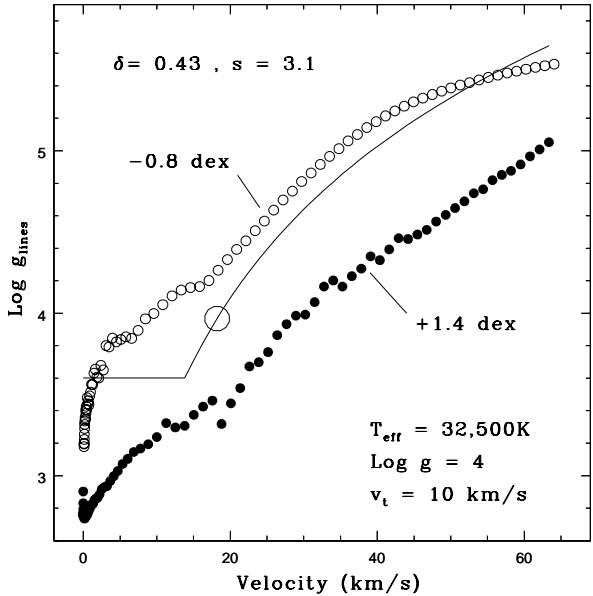}
\caption{Departures from dynamical consistency. The solid line
shows the variation of $g_{\ell} (v)$ assumed in deriving the stratification
of the RL with the indicated parameters, for which 
$J_{L}(gm \: cm^{-2} s^{-1})$ = -7.32 dex - see Fig.3. 
The filled circles are the MC estimates $\tilde{g_{\ell}}$ when $J$ is
increased to -5.92 dex; and the small open circles are the estimates for
$J$ = -8.12 dex. The large open circle is the sonic point.}
\end{figure}

\subsection{Two strong-wind stars}

Given the large discrepancies for the weak-wind stars between $J_{M}, J_{L}$
and $J_{V}$, a reader might reasonably conclude that, since no two methods
even approximately agree, there is no basis for preferring one estimate over
another. The data in Table 1 is therefore now used to
compute $J$'s for the two Galactic O4 stars investigated by Bouret et al.
(2005). These authors are co-authors of the Marcolino
paper, and the same diagnostic codes, TLUSTY and CMFGEN, are used.

For the O4 V star HD 96715, Bouret et al derive   
$\Phi = 2.5 \pm 0.5 \times 10^{-7} \cal{M}_{\sun}$$ yr^{-1}$, 
corresponding to
$log J = -5.74 \pm 0.08$. This is their preferred estimate, derived with
the filling factor of clumps as an additional adjustable parameter. For
a homogeneous wind, $\Phi$ is 0.86 dex higher, corresponding to
$log J = -4.88$. 

The atmospheric parameters derived by Bouret et al. are 
$T_{eff} = 43500 \pm 2200K$ and $log g = 4.0 \pm 0.1$. Interpolation
in Table 1 gives $log J_{L} =  -5.54 \pm 0.25$. But note that Table 1
is computed with $v_{t} = 10 km s^{-1}$, whereas Bouret et al derive
$v_{t} = 15 \pm 5 km s^{-1}$. According to Table 1 in L07a, this increase in $v_{t}$
reduces $J_{L}$ by about 0.30 dex, bringing the prediction to
$log J_{L} =  -5.84 \pm 0.36$. This agrees within errors with the clump-
corrected value of Bouret et al. but is in strong disagreement with the value
for a homogeneous wind.

The second Bouret et al. star is the O4 I supergiant HD 190429A. The
clumped-wind analysis gives
$\Phi = 1.8 \pm 0.4 \times 10^{-6} \cal{M}_{\sun}$$ yr^{-1}$ ,
corresponding to $log J = -5.31 \pm 0.08$. The atmospheric parameters
derived by Bouret et al. are   
$T_{eff} = 39000 \pm 2000K$ and $log g = 3.6 \pm 0.1$ giving
$log J_{L} =  -5.04 \pm 0.76$, where the large standard error is due to the 
steep gradient of $J_{L}$ at this star's location in  $(T_{eff},g)$-space. 
As before, we now apply a correction of -0.30 dex because the measured 
$v_{t} = 15 \pm 2 km s^{-1}$. The final prediction is therefore 
$log J_{L} =  -5.34 \pm 0.77$, consistent with the clumped wind
analysis but, in this case, also with the value $log J = -4.78 \pm 0.08$ 
for a homogeneous wind.

The clumped-wind $\Phi$'s derived by Bouret et al are lower than the Vink
et al. predictions by 0.9 dex for HD 96715 and 0.5 dex for HD 190429A.
On this basis, Bouret et al. conclude that our understanding of O-star winds
requires fundamental revision. The success here in reproducing their
estimates shows that the required revision is simply the
inclusion of a dynamically consistent transition to the observed photosphere.  
In addition, this theoretical confirmation of their low rates
reinforces their conclusion that the evolutionary tracks of massive stars
need to be recomputed.

\subsection{Mechanism}

The internal details of the RL models can be investigated in order to
understand the sharp rise in $J$ as one moves from late- to early-
type O dwarfs. Most illuminating are the fractional contributions of
various ions to $\tilde{g_{\ell}}$ in the layer containing the sonic point. 
Such calculations have been carried out for the models with $log \: g = 4$.
These reveal that the high $J$'s for $T_{eff} \ga 40,000K$ are
due to the ability of high ions to
absorb momentum from the emergent flux below the Lyman limit, as has
long been
understood (Castor, Abbott, \& Klein 1976; Lamers \& Morton 1976). Of
dominant importance is the changing ionization balance of iron, resulting
in the effective driving ion Fe V replacing the ineffective ion 
Fe IV (cf. Vink et al. 2000). Thus, in the weak-wind domain at
$T_{eff} = 32,500K$ , Fe V lines contribute only 0.7\% of 
$\tilde{g_{\ell}}$ at
$v = a$, but this rises to 44\% at 40,000K and reaches a maximum of 78\% at   
45,000K.

These results show that the termination of strong winds in $(T_{eff},g)$-space 
is determined by ionization balance in the {\em reversing layer}.
In these models (L07a, Sect.2.5), this balance
is matched at a representative point to that of the corresponding static TLUSTY
atmosphere.

\section{Conclusion}

The aim of this paper has been to respond to an apparent problem for the
theory of radiatively-driven winds arising from the extraordinarily low
$\Phi$'s found for late-type Galactic O stars, as exemplified recently by M09.
The problem is lessened,
though perhaps not entirely resolved, when the observed rates are compared to
predictions made from first principles using the theory of dynamic RLs 
(LS70; L07a) rather than to the semi-empirical estimates of Vink et al. (2000).
When this theory is used to define a weak-wind domain in $(T_{eff},g)$-space, 
the M09 stars are found to be within or at the edge of this domain. It remains
to be seen whether future developments in our understanding
of low density supersonic winds will close the remaining gap between predicted
and observed rates. 

A still outstanding issue is the huge disparity between the $\Phi_{V}$'s and
the
observed estimates. It would be informative if the Vink et al. code were used
to fit the M09 stars individually and the emergent MC spectra compared
with the observed spectra, as Abbott \& Lucy (1985) did for $\zeta$ Puppis. 
From experiments reported by M09, this consistency check is expected to
fail, revealing that the $\Phi_{V}$'s give rise to unacceptably strong wind 
lines. Such discrepancies would then provide the basis for revising the 
Vink et al assumptions about physical conditions at velocities 
$\sim 0.5 v_{\infty}$ in these stars' winds.

This substantial progress towards resolving the {\em weak-wind problem}
and the success in reproducing the clumped-wind $\Phi$'s 
of Bouret et al. (2005) strongly supports the simple picture  
for the onset of winds presented in LS70: selective radiation pressure expels
high atmospheric layers causing a pressure inbalance in the photosphere 
resulting in an upwelling of matter from deeper layers. The natural 
steady-state configuration is then the laminar transonic flow described
by the theory of moving RL's (LS07; L07a), which merits still further 
development. 

\acknowledgement

I am grateful to J.S.Vink, A. de Koter and T.L.Hoffmann for prompt answers
to queries relevant to the refereeing of this paper.

\end{document}